\documentclass[conference]{IEEEtran}

\RequirePackage[american]{babel}
\IEEEoverridecommandlockouts

\usepackage[square,numbers]{natbib}
\usepackage{amsmath,amssymb,amsfonts}
\usepackage{algorithmic}
\usepackage{graphicx}
\usepackage{textcomp}
\usepackage{xcolor}

\usepackage{tikz,url}

\usepackage{filecontents}

\usepackage{stmaryrd}
\usepackage{breqn}
\usepackage[ruled,linesnumbered]{algorithm2e}
\usepackage{siunitx}
\usepackage{tikz} 
\usepackage{pgfplots}
\usepackage[inline]{enumitem}

\usepackage{multirow}
\usepackage{textcomp}
\usepackage{adjustbox}
\usepackage{mathrsfs}
\usepackage{subfig}
\usepackage{svg}
\usepackage{comment}

\usetikzlibrary{patterns}
\usepgfplotslibrary{groupplots}

\usepgfplotslibrary{units} 
\usepgfplotslibrary{colorbrewer}
\pgfplotsset{width=14cm,compat=1.9}

\newcommand{\etal}{\mbox{\emph{et al.\ }}}
\newcommand{\ie}{\mbox{i.e.}} 
\newcommand{\eg}{\mbox{e.g.}} 
 
\newcommand{\st}{\mbox{\emph{s. t.\ }}} 
\newcommand{\AndGate}[2]{${#1} 	\land {#2}$}
\newcommand{\XorGate}[2]{${#1} \mathbin{\oplus} {#2}$}

\newcommand{\Oplus}{\ensuremath{\vcenter{\hbox{\scalebox{1}{$\oplus$}}}}}

\newcommand*\fourteenpt{\fontsize{14}{15.5}\selectfont}

\newcommand{\cparallel}{{\mkern2mu\vphantom{\perp}\vrule depth 0.5pt\mkern5mu\vrule depth 0.5pt\mkern3mu}}

\begin{document}

\date{}

\title{\fourteenpt \bf CPU and GPU Accelerated Fully Homomorphic Encryption\textsuperscript{**}\thanks{** Accepted in  2020 IEEE International Symposium on Hardware Oriented Security and Trust (HOST)}}

\author{Toufique Morshed\IEEEauthorrefmark{1},  \IEEEauthorblockN{Md Momin Al Aziz\IEEEauthorrefmark{2} and Noman Mohammed\IEEEauthorrefmark{3}}
\IEEEauthorblockA{Computer Science,
University of  Manitoba\\
Email: \IEEEauthorrefmark{1}morshed@cs.umanitoba.ca, \IEEEauthorrefmark{2}azizmma@cs.umanitoba.ca,
\IEEEauthorrefmark{3}noman@cs.umanitoba.ca}}

\maketitle
\begin{abstract}
Fully Homomorphic Encryption (FHE) is one of the most promising technologies for privacy protection as it allows an arbitrary number of function computations over encrypted data. However, the computational cost of these FHE systems limits their widespread applications. In this paper, our objective is to improve the performance of FHE schemes by designing efficient parallel frameworks. In particular, we choose Torus Fully Homomorphic Encryption (TFHE)~\cite{chillotti2016faster} as it offers exact results for an infinite number of boolean gate (\eg, AND, XOR) evaluations. 
We first extend the gate operations to algebraic circuits such as addition, multiplication, and their vector and matrix equivalents. Secondly, we consider the multi-core CPUs to improve the efficiency of both the gate and the arithmetic operations.

Finally, we port the TFHE to the Graphics Processing Units (GPU) and device novel optimizations for boolean and arithmetic circuits employing the multitude of cores.
We also experimentally analyze both the CPU and GPU parallel frameworks for different numeric representations (16 to 32-bit). Our GPU implementation outperforms the existing technique~\cite{chillotti2016faster}, and it achieves a speedup of $20\times$ for any 32-bit boolean operation and $14.5\times$ for multiplications.
\end{abstract}

\section{Introduction}\label{sec:introduction}
    Fully Homomorphic Encryption (FHE)~\cite{gentry2009fully}
    has attracted attention in modern cryptography research. FHE cryptosystems provide strong security guarantee and can compute an infinite number of operations on the encrypted data. Due to the emergence of various data-oriented applications \cite{pham2017oride,kim2017secure,chen2018logistic} on sensitive data, the idea of computing under encryption has recently gained momentum. FHE is the ideal cryptographic tool that addresses this privacy concern by enabling computation on encrypted data.
    
    \noindent\textbf{Motivating Applications.} There has been significant advancements in machine learning techniques and their applications over the last few years. The usage and accuracy of such methods have surpassed the state of the art solutions in manifolds. We can attribute three components behind this improvement: a) better algorithms, b) big data and c) efficient hardware  (H/W) enabled parallelism. With the increase of cloud services, several service providers (\eg, Google Prediction API~\cite{googlepredictionAPI}, Azure Machine Learning~\cite{azure}) have combined the three attributes to facilitate machine learning as a service. 
     
    In these services, users outsource their data to the cloud server to build a machine learning model. However, data outsourcing exposes the sensitive data to the cloud service provider \cite{sadat2019privacy} and thus susceptible to privacy attacks by the employees at the service provider \cite{uber_employee}. FHE schemes are practical for such use cases as these schemes facilitate computation on encrypted data. Using FHE, a data owner can encrypt the sensitive data before outsourcing it to the server, and also the server can execute the required machine learning algorithm for data analysis. 
    
    \subsection{Current Techniques}
        \begin{table*}[t]

\centering
\caption{A comparative analysis of existing Homomorphic Encryption schemes for different parameters on 32-bit number.}
\label{tbl:comparativeAna}
\resizebox{\linewidth}{!}{
\begingroup
\begin{tabular}{l|c|c|c|c|c|c|c|c}
\hline
                                                          & \textbf{Year}   & \textbf{Homomorphism}   & \textbf{Bootstrapping}         & \textbf{Parallelism}                                             & \textbf{Bit security} & \textbf{Size (kb)} & \textbf{Add. (ms)}         & \textbf{Mult. (ms)}  \\ \hline 
RSA~\cite{rivest1978method}         & 1978   & Partial & $\times$ & $\times$                                   &    128          &      0.9    & $\times$ &          5            \\ \hline
Paillier~\cite{paillier1999public}  & 1999   & Partial       & $\times$ & $\times$                                   &  128            &  0.3 &   4 & $\times$ \\ \hline
TFHE~\cite{chillotti2016faster}     & 2016   & Fully          & Exact                 & AVX~\cite{lomont2011introduction} & 110    & 31.5      & 7044              & 4,89,938            \\ \hline
HEEAN~\cite{cheon2018bootstrapping} & 2018   & Somewhat       & Approximate           & CPU                                                     & 157    & 7,168     & 11.37                 & 1,215             \\ \hline
SEAL (BFV)~\cite{sealcrypto}              & 2019   & Somewhat       & $\times$ & $\times$                                   & 157    & 8,806   & 4,237                 & 23,954               \\ \hline
cuFHE~\cite{cuFHE}                                                & 2018 & Fully          & Exact                 & GPU                                                     & 110    & 31.5      & 2,032
              & 1,32,231            \\ \hline
NuFHE~\cite{NuFHE}                                                & 2018 & Fully          & Exact                 & GPU                                                     & 110    & 31.5      & 4,162
              & 1,86,011            \\ \hline
Cigulata~\cite{cingulata}                                                & 2018 & Fully          & Exact                 & $\times$                                                     & 110    & 31.5      & 2,160
              & 50,690            \\ \hline

Our Method & - & Fully          & Exact                 & GPU                                                     & 110    & 31.5      & 1,991
              & 33,930            \\ \hline
\end{tabular}
\endgroup}
\color{black}
\end{table*} 
        
        \textcolor{black}{The homomorphic encryption schemes can be divided into three major categories: Partially, Somewhat, and FHE schemes. Partially Homomorphic schemes only support one type of operation (e.g., addition or multiplication); such schemes are not useful in performing arbitrary computations on encrypted data.}
    
        Somewhat Homomorphic Encryption (SWHE) schemes are more equipped than partially homomorphic encryption schemes. These schemes support both addition and multiplication operations on encrypted data, but for a limited (or pre-defined) number of times. In addition, these schemes are relatively efficient (see Table~\ref{tbl:comparativeAna} for comparison) and therefore are practical for certain applications. However, even these schemes require complex parameterization and are not powerful enough for more complicated operations such as deep learning.
    
        FHE schemes support both addition and multiplication operations for an arbitrary number of times. This property allows computing any function on the encrypted data. Both SWHE and FHE use the Learning with Error (LWE) paradigm, where an error is introduced with the ciphertext value to guarantee security~\cite{regev2009lattices}. This error grows with each operation (especially multiplication) and causes incorrect decryption after a certain number of operations. Therefore, this error needs to be minimized to support arbitrary computation. The process of reducing the error is called Bootstrapping. FHE employs bootstrapping after a certain number of operations resulting in higher computation overhead, while SWHE provides faster execution time by limiting/pre-defining the number of operations on the encrypted data.
        
        The above discussion provides an intuition about the applications of different HE schemes. That is, SWHE is better suited for the applications where the computational depth is shallow and known (/fixed) prior to the computations. However, these schemes are not suitable for applications that require arbitrary depth like deep learning. In order to compute complicated functions like deep learning, the researchers have proposed alternative models that require the existence of a third party~\cite{sadat2019safety, juvekar2018gazelle, hesamifard2018privacy}. The aim is to minimize the propagated error without executing the costly bootstrapping procedure for SWHE schemes. However, such an assumption (i.e., the existence of a trusted third party) is not always easy to fulfill.  In this article,  we assume that the computational entity (e.g., cloud server) is standalone, and we show that parallelism can be used to lower the cost of FHE instead of relaxing the security assumptions for the computation model.
    
        \noindent\textbf{Why TFHE?}
        \textcolor{black}{There have been several attempts in improving the asymptotic performance and numerical operations of FHE \cite{brakerski2013packed, ducas2014,fan2012somewhat}, which are pivotal to this work (Section~\ref{sec:relWorls} for details)}. Torus FHE (TFHE)~\cite{chillotti2016faster} is one of the most renowned FHE schemes that meets the expectation of arbitrary depth of circuits with faster bootstrapping technique. TFHE  also incurs lower storage requirement compared to the other encryption schemes (Table~\ref{tbl:comparativeAna}). The plaintext message space is binary in TFHE. Hence, the computations are based solely  on boolean gates, and each gate operation entails a bootstrapping procedure in gate bootstrapping mode.
    
        \noindent\textbf{Why GPU?} \textcolor{black}{Most of the FHE schemes are based on the Learning With Errors (LWE), where plaintexts are encrypted using polynomials and can be represented with vectors. Therefore, most computations are operated on vectors that are highly parallelizable. On the contrary, Graphics Processing Units (GPUs) offer a large number of computing cores (compared to CPUs). These cores can be utilized to compute parallel vectors operations. Therefore, we can utilize these cores to parallel the FHE computations. However, we also have to consider the fixed and limited memory of GPUs (8-16GB) and their reduced computing power compared to any CPU core.}

        \subsection{Contributions} 
            \begin{table}[t]
\centering
\caption{A comparison of the execution times (sec) of TFHE \cite{chillotti2017faster} and our CPU, GPU  framework for 32-bit numbers }
\label{tab:contribution}
\resizebox{\linewidth}{!}{
\begingroup
\begin{tabular}{l|c|c|c|c|c}
\hline
\multirow{2}{*}{} & \multirow{2}{*}{\textbf{Gate Op.}} & \multicolumn{2}{c|}{\textbf{Addition}} & \multicolumn{2}{c}{\textbf{Multiplication}} \\ \cline{3-6} 
                  &                           & \textbf{Regular}        & \textbf{Vector}       & \textbf{Regular}          & \textbf{Matrix (mins)}           \\ \hline
TFHE~\cite{chillotti2016faster}              & 1.40                       & 7.04           & 224.31       & 489.93           & 8,717.89          \\ \hline
CPU-Parallel  & 0.50                       & 7.04           & 77.18        & 174.54           & 2,514.34          \\ \hline
GPU-Parallel  & 0.07                       & 1.99           & 11.22        & 33.93            & 186.23           \\ \hline
\end{tabular}
\endgroup}
\end{table}
            \begin{itemize}
                \item In this paper, we extended the boolean gate operations (AND, XOR, etc.), from earlier work \cite{chillotti2016faster}, to higher level algebraic circuits (\eg, addition, multiplications).
                
                \item Initially, we constructed a CPU parallel TFHE framework  as a baseline to leverage the available computational cores. Experimental results demonstrate the advantage of such construction using multi-threading as they outperform the sequential implementation.
                \item We devised boolean gates using the GPU parallelism, and employed novel optimizations such as bit coalescing, compound gates, and tree-based additions to implement the higher level algebraic circuits. We also modified and incorporated parallelism to the existing algorithms: Karatsuba and Cannon for multiplication and matrix counterpart to achieve further speedup. The code is readily available at \url{https://github.com/toufique-morshed/CPU-GPU-TFHE}.
                
                \item 
                    We have conducted several experiments to compare the computation time of the sequential TFHE~\cite{chillotti2016faster} with our proposed CPU and GPU parallel frameworks.  We have also outlined a real-world application with Linear Regression on different datasets ~\cite{sadat2018secure}. From Table~\ref{tab:contribution}, our proposed GPU $\cparallel{}$ method is $14.4\times$ and $46.81\times$ faster than the existing technique for regular and matrix multiplications, respectively. We have also benchmarked our performance with existing TFHE frameworks on GPU, namely, cuFHE~\cite{cuFHE}, NuFHE~\cite{NuFHE}.
            \end{itemize}
            Notably, the existing GPU enabled TFHE libraries, cuFHE~\cite{cuFHE} and NuFHE~\cite{NuFHE}, have implemented the TFHE boolean gates using GPUs, whereas our goal was to construct an optimized arithmetic circuit framework. Our design choices and algorithms reflect this improvement and resultantly, our multiplications are around $3.9$ and $4.5$ times faster than cuFHE and NuFHE,  respectively.

\section{Preliminaries}\label{sec:preliminaries}
    \begin{table}[t]
    \centering
    \caption{Notations 
    used throughout the paper}\label{tbl:notation}
    \begin{adjustbox}{max width=\columnwidth}  
    \begingroup
        \begin{tabular}{c|l}
            \hline
            \textbf{Notation}            &   \textbf{Description} \\ \hline
            $\textbf{A} \in \mathbb{Z}^{r\times c}$       &   Integer matrix of dimension $r\times c$ \\ \hline
            $\overrightarrow{A} \in \mathbb{Z}^{\ell}$  &   Integer vector of length $\ell$\\ \hline
            $A, A' \in \mathbb{Z} \subset \mathbb{B}^{n}$         &   $n$-bit integer number and its complement\\ \hline
            $a_i \in \mathbb{B}$                               &   Binary bit at position $i$\\ \hline
            $\mathbb{L}^{n}$                                                &   LWE sample vector of length $n$\\ \hline
            $n$, $m$                                                             &   Bit size and Secret key size\\ \hline
            $\cparallel$, $\ll$                                                    &   Parallel and Left Shift operation\\ \hline
            \AndGate{}{}, $|$, \XorGate{}{}                                 &   Binary AND, OR and XOR operation\\ \hline
        \end{tabular}
        \endgroup
    \end{adjustbox}
\end{table}

    \subsection{Torus FHE (TFHE)}\label{sec:backgroung:tfhe}
        
        \noindent In this work, we closely investigate Torus FHE (TFHE) \cite{chillotti2016faster} where the plain and ciphertexts are defined over a real torus $\mathbb{T} = \mathbb{R}/\mathbb{Z}$, a set of real numbers modulo 1. The ciphertexts are constructed  over Learning with Errors (LWE)~\cite{regev2009lattices} and represented as Torus LWE (TLWE) where an error term (sampled over a Gaussian distribution $\chi$) is added to each ciphertext. For a given dimension $m \geq 1$ (key size), secret key $\overrightarrow{S}\in \mathbb{B}^{m}$ ($m$-bit binary vector), and error $e\in \chi$, an LWE sample is defined as ($\overrightarrow{A}, B$) where $\overrightarrow{A}\in \mathbb{T}^m$ \st $\overrightarrow{A}$ is a vector of torus coefficients of length $m$ (key size) and each element $A_i$ is drawn from the uniform distribution over $\mathbb{T}$, and 
            $B = \overrightarrow{A} \cdot \overrightarrow{S} + e$.
        
        \label{sec:bootstrapping}
        The error term ($e$) in LWE sample grows and propagates with the number of computations (\eg, addition, multiplication). Therefore, bootstrapping is introduced to decrypt and re-encrypt the ciphertexts under encryption to remove the noise.

        TFHE considers the binary bits as plaintext and generates LWE samples as ciphertexts. Hence, LWE sample computations in ciphertext are analogous to binary bit computations in plaintext. As a binary vector represents an integer number, an LWE sample vector ($\mathbb{L}^{n}$) can represent an encrypted integer. 
        For example, an $n$-bit integer becomes $n$-LWE sample after encryption. Thus, the boolean gate operations of an addition circuit between two $n$-bit numbers correspond to the similar operations on LWE samples of encrypted numbers. Throughout this paper, we use \emph{bit} and \emph{LWE sample} interchangeably. Here, we choose TFHE for the following reasons:
        \begin{itemize}
            \item \textbf{Fast and Exact Bootstrappping.}
                TFHE provided the fastest and exact bootstrapping technique which only required around $0.1 s$. Some recent encryption schemes \cite{cheon2018bootstrapping,boura2018chimera} also proposed faster bootstrapping and FHE computations in general.  However, they do not perform exact bootstrapping and erroneous after successive computations on the same ciphertexts.
            \item \textbf{Ciphertext Size.}
            Compared to the other HE schemes, TFHE offers smaller ciphertext size as it operates on binary plaintexts as shown in Table~\ref{tbl:comparativeAna}. Nevertheless, this minimal storage advantage allowed us to utilize the limited and fixed memory of GPU when we optimize the gate structures.
            \item \textbf{Boolean Operations.}
                TFHE also supports boolean operations which are extended to construct arbitrary functions. These binary bits can then be operated in parallel if their computations are independent of each other.
        \end{itemize} 
     \color{black}
    
        \noindent\textit{Existing Implementation:}
            The current TFHE implementation comes with the basic cryptographic functions (\ie, encryption, decryption, etc.) and all binary gate operations. Although the gates are computed somewhat sequentially in the original implementation \cite{chillotti2016faster}, the underlying architecture uses Advanced Vector Extensions (AVX)~\cite{lomont2011introduction}. AVX 
            is an extension to x$86$ instruction set from Intel which facilitates parallel vector operations. The bootstrapping  procedure requires expensive Fast Fourier Transform (FFT) operations ($O(n\log n)$). The existing implementation uses the Fastest Fourier Transform in the West (FFTW)~\cite{frigo1998fftw} which inherently uses AVX.
        
    \subsection{Parallelism}\label{sec:backgroung:parallelism}
        
        Our \noindent\textbf{CPU Parallel} framework utilizes the CPU cores and existing resources (\ie, Vector Extensions) available in a typical computing machines. We exploit these parallel components on  CPUs as we breakdown each algebraic computation into independent parts and distribute them among the available resources. However, one major drawback of the CPU framework is the limited number of available cores. Contemporary desktop computers come with 4 to 8 cores containing a maximum of 16 threads. There have been multiple attempts~\cite{salem2019privacy} to use a large number of CPUs collectively for parallel operations, whereas we show that single  GPU is equivalent (and better performing) for most FHE operations.
        
        \textcolor{black}{In contrast to CPUs, \textbf{GPU Parallel}  framework offers a significant number of cores available solely from the hardware. Thus, the expense of increasing cores with multiple CPUs is reduced by integrating one GPU. Nevertheless, we had to consider the two major shortcomings which are: a) limited global memory and b) communication time through PCIe.}


\section{Sequential Framework}\label{sec:circuitConstruction}
    \subsection{Addition}\label{sec:addition}
    
        A carry-ahead 1-bit full adder circuit takes two input bits along with a carry to compute the sum and a new carry that propagates to the next bit's addition. Therefore, in a full adder, we have  three inputs as $a_i, b_i$ and $c_{i-1}$, where $i$ denotes the bit position. Here, the addition of bit $a_1$ and  $b_1$ in $A, B \in \mathbb{B}^{n}$ requires the carry bit from $a_0$ and  $b_0$. This dependency enforces the addition operation to be sequential for $n$-bit numbers~\cite{catherine1993paralleladdition}.
        
        \color{black}
    \subsection{Multiplication}\label{sec:multiplication}
    \subsubsection{Naive Approach}\label{sec:multiplicationNaive}
        For two $n$-bit numbers $A,B \in \mathbb{Z}$, 
        we multiply (\texttt{AND}) the number $A$ with each bit $b_i\in B$,  resulting in $n$ numbers. Then, these numbers are \texttt{left shift}ed by $i$ bits individually resulting in $[n,2n]$-bit numbers. Finally, we \texttt{accumulate} (reduce by addition) the $n$ shifted numbers using the addition. 
    \subsubsection{Karatsuba Algorithm}\label{sec:multiplicationKaratsuba}
        We consider the divide-and-conquer Karatsuba's algorithm for its improved time complexity  $O(n^{\log^{3}})$~\cite{karatsuba1962multiplication}. It relies on dividing the large input numbers and performing smaller multiplications. For $n$-bit inputs, Karatsuba's algorithm splits them into smaller numbers of $n/2$-bit size and replaces the multiplication by additions and subsequent multiplications (Line \ref{ln:karat:changeEnd} of Algorithm~\ref{algo:karatsuba}). Later, we introduce parallel vector operations for further optimizations.

\begin{algorithm}[t]
    \DontPrintSemicolon
    \caption{Karatsuba Multiplication \cite{karatsuba1962multiplication}}\label{algo:karatsuba}
    \KwIn{X, Y $\in \mathbb{B}^n$}
    \KwOut{Z $\in \mathbb{B}^{2n}$}
    \SetKwProg{}{}{}{}
    \If{$n <n_{0}$}{
        \Return{} \FuncSty{BaseMultiplication(X, Y)}\;
    }
    X$_{0}\leftarrow$ X$\mod 2^{n/2}$\;
    Y$_{0}\leftarrow$ Y$\mod 2^{n/2}$\;
    X$_{1}\leftarrow$ X/2$^{n/2}$\;
    Y$_{1}\leftarrow$ Y/2$^{n/2}$\;
    
    Z$_{0} \leftarrow \FuncSty{KaratsubaMultiply}$(X$_0$, Y$_0$)\;\label{ln:karat:changeStart}
    Z$_{1} \leftarrow \FuncSty{KaratsubaMultiply}$(X$_1$, Y$_1$)\;
    Z$_{2} \leftarrow \FuncSty{KaratsubaMultiply}$(X$_0$ + Y$_0$, X$_1$ + Y$_1$)\;
    
    \Return{Z $\leftarrow$ Z$_0$ + (Z$_2$ - Z$_1$ - Z$_0$) $2^{n}$ + (Z$_1$)$2^{2n}$}\label{ln:karat:changeEnd}
\end{algorithm}

\section{CPU-based Parallel Framework }\label{sec:openMPFra}
    We propose a CPU $\cparallel{}$ framework utilizing the multiple cores available in computers. Since the existing TFHE implementation uses AVX2, we employ that in our CPU $\cparallel{}$ framework.
    
    \subsection{Addition}\label{sec:openMPFra:Addition}
    \begin{figure}[t]
            \centering
            \includegraphics[width=0.85\linewidth,scale=1]{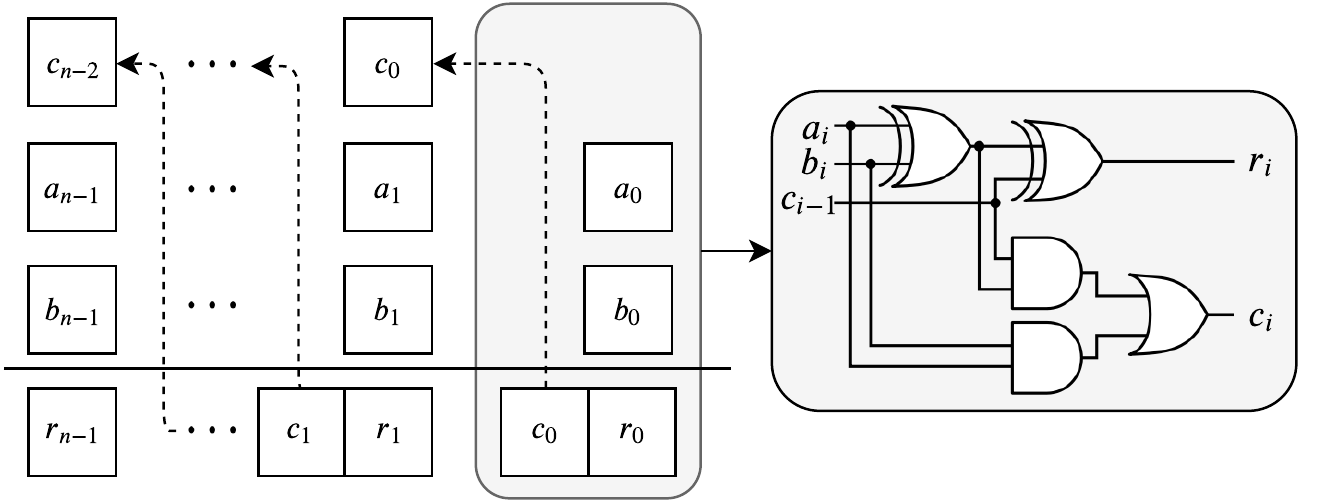}
            \caption{Bitwise addition of two $n$-bit numbers $A$ and $B$. 
            $a_i, b_i, c_i, r_i$ are $i^{th}$-bit of $A,B$, $carry$, and  the $result$}
            \label{fig:bitAddition}
        \end{figure}
        Figure~\ref{fig:bitAddition} illustrates the bitwise addition operation considered in our CPU framework. Here, any resultant bit $r_i$ depends on its previous $c_{i-1}$ bit. 
        The dependency restricts incorporating any data-level parallelism in the addition circuit construction.
        
        Here, it is possible to exploit task-level parallelism where two 
        threads execute the \texttt{XOR} 
        and \texttt{AND} 
        operations (Figure~\ref{fig:bitAddition}), simultaneously. We observed that the time required to perform such \texttt{fork-and-join} between two threads is higher than executing them serially. This is partially due to the costly thread operations and eventual serial dependency of the results. Hence, we did not employ this technique for CPUs.
        
        
    \subsection{Multiplication}\label{sec:openMPFra:Multiplication}
        Out of the three major operations (\texttt{AND}, \texttt{left shift}, and \texttt{accumulation} in multiplications,
        the \texttt{AND} and left shifts 
        can be executed in parallel. For example, for any two $16$-bit numbers A, B ($\in \mathbb{B}^{16}$), and four available threads, we divide the \texttt{AND} and  \texttt{left shift} operation among four threads. 
        
        
        On the other hand, the accumulation operation is demanding as it requires $n$ (for $n$-bit multiplication) additions. The accumulation operation adds and stores values to the same variable, which makes it atomic. Therefore, all threads performing the previous \texttt{AND} and \texttt{left shift} have to wait for such accumulation which is termed as global thread synchronization~\cite{chandra2001parallel}. Given that it is computationally expensive, we do not employ this technique in any parallel framework.
        
        We utilized a custom reduction operation in OpenMP~\cite{chandra2001parallel}, which uses the global shared memory (CPU) to store the in-between results. This customized reduction foresees additions of any results upon completion and facilitates a performance gain by  avoiding the global thread synchronization. 
        
        
    \subsection{Vector Operations}\label{sec:openMPFra:Vector}
        
        To compute the vector operations (addition, multiplication) efficiently, we distribute the work into multiple threads.
        For example, $\overrightarrow{A}$, $\overrightarrow{B}$, and $\overrightarrow{C} \in \mathbb{Z}^{\ell}$ are three vectors of length $\ell$, where $\overrightarrow{C}=\overrightarrow{A}+\overrightarrow{B}$, and each element A$_{i}$, B$_{i}$, or C$_{i}$ are $n$-bit integers. The computation of each position of $\overrightarrow{C}$ is independent. Hence, 
        we can share the work among different threads. We take similar measures for multiplication as well.
    
    \subsection{Matrix Operations}\label{sec:openMPFra:matrix}
        
        \noindent\textit{Matrix Addition}\label{sec:openMPFra:matrix:addition}
            is a series of addition operations between the elements of
            two matrices. 
            The addition operations between them are independent of each other. Therefore, we divide the matrices row-wise and distribute the additions among threads.
        
        \noindent\textit{Matrix Multiplication}\label{sec:openMPFra:matrix:multiplication}
            is a bit more complicated than addition. It consists of both multiplications and additions.
            
            \begin{align*}
                \textbf{X}\times \textbf{Y}
                &=\begin{bmatrix}
                    X_{00}Y_{00} + X_{01}Y_{10} & X_{00}Y_{01} + X_{01}Y_{11}\\
                    X_{10}Y_{00} + X_{11}Y_{10} & X_{10}Y_{00} + X_{11}Y_{10}\\
                \end{bmatrix} 
            \end{align*}

            
            Inspecting the calculations above, in a 2x2 matrix multiplication, we highlight three major observations.
            \begin{itemize}
                \item Each element of the resultant matrix is independent.
                \item All multiplication operations are independent.
                \item The addition operation is accumulation operation.
            \end{itemize}

            \noindent Here, the computation for rank $2\times 2$ end with an addition operation. However, for rank $3$ ($>2$), the computation for the first index becomes $[\textbf{X}\times \textbf{Y}]_{00} = X_{00}Y_{00} + X_{01}Y_{10} + X_{02}Y_{20}$, where all the multiplication results are accumulated (reduced by addition). We address the incorporation of parallelism in such accumulation in Section~\ref{sec:gpuPar:Multiplication}.
            
            
            The small and limited number of cores is a limitation while distributing such matrix operation for CPU $\cparallel{}$. Hence, we only take the first observation into account and employ each core to compute the results  for the CPU-based parallel framework. 
           

\section{GPU-based Parallel Framework }\label{sec:gpuPar}
    
    In this section, we first present three generalized techniques to introduce GPU parallelism (GPU $\cparallel$) for any FHE computations. Then, we adopt them to implement and optimize the arithmetic operations.
    
    \subsection{Proposed Techniques}
    
        \subsubsection{Parallel TFHE Construction}\label{sec:gpuPar:TfheCons}
            
\begin{figure}[t]
                \centering
                \includegraphics[width=0.78\linewidth]{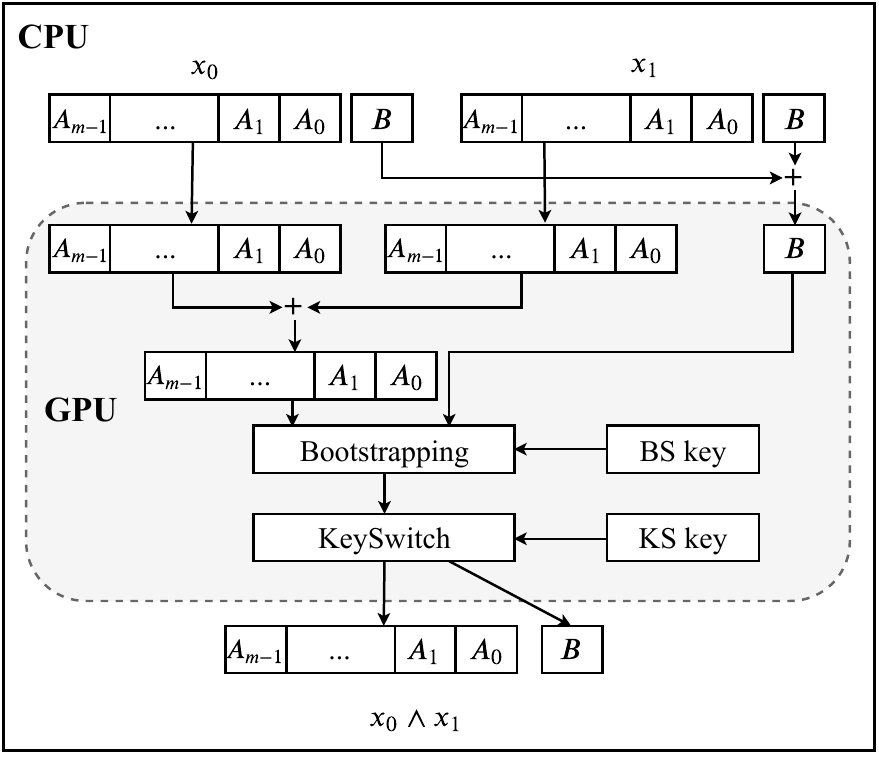}
                \caption{Arbitrary operation between two bits where  BS, KS key represents bootstrapping and key switching keys, respectively}
                \label{fig:gpu_execution}
            \end{figure}

            \color{black}
            We depict the boolean circuit computation in Figure~\ref{fig:gpu_execution}. Here, each LWE sample comprises of two variables namely $\overrightarrow{A}$ and $B$, where $\overrightarrow{A}$ is defined as a vector. It is noteworthy that $\overrightarrow{A}$ is a 32-bit integer vector defined by the secret key size ($m$) which has a lower memory requirement compared to  other FHE implementations (Section~\ref{sec:relWorls}). In our parallel TFHE construction, we only store the vector $\overrightarrow{A}$ on the GPU's global memory.

            In addition to all vector operations inside the GPU, we also employ the native cuda enabled FFT library (cuFFT) which uses the parallel cuda cores for FFT operations. Here, the parallel batching technique from cuFFT supports multiple FFT operations to be executed simultaneously. However, cuFFT also limits  such parallel number of batches. It keeps the batches in an asynchronous launch queue, and processes a certain number of batches in parallel. This number of parallel batches solely depends on the hardware capacity and specifications~\cite{gpuspec}.
            
        \subsubsection{Bit Coalescing (BC)}\label{sec:gpuPar:BitCoal}
\begin{figure}[t]
                \centering
                \includegraphics[width=0.75\linewidth]{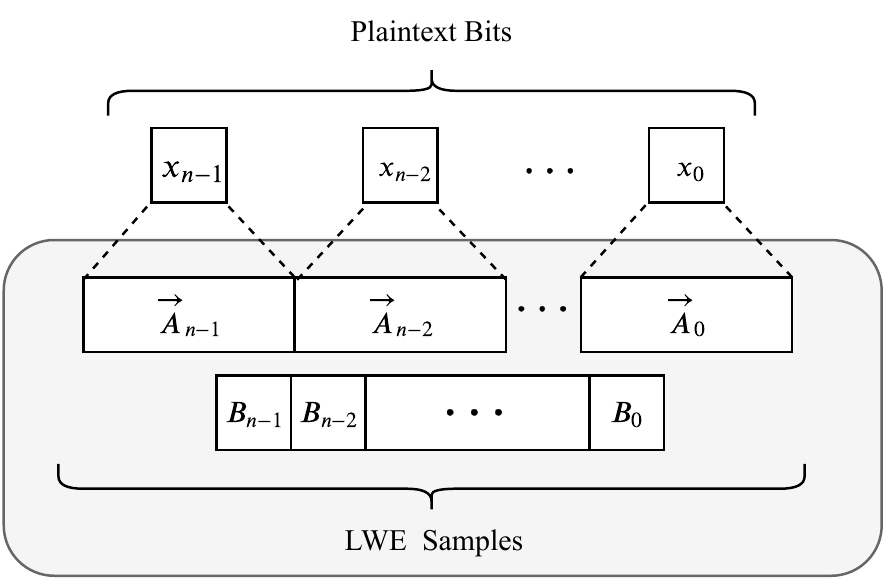}
                \caption{Coalescing $n$-LWE samples (ciphertexts) for $n$-bits}
                \label{fig:bitcoal}
            \end{figure}

            Bit Coalescing combines $n$-LWE samples in a contiguous memory to represent $n$-encrypted bits. The encryption of a $n$-bit number, $X \in \mathbb{B}^n$ requires $n$-LWE samples (ciphertext), and each sample contains a vector of length $m$. Instead of treating the vectors of ciphertexts separately, we coalesce them altogether (dimension $1\times mn$) as illustrated in Figure~\ref{fig:bitcoal}.
            
            The intuition behind such construction is to increase parallelism by extending the vector length in a contiguous memory. Coalescing the vectors increases the vector length but we incorporate more threads to maximize parallelization and reduce the execution time.

        \subsubsection{Compound Gate}\label{sec:gpuPar:compound}
        Since addition is used in most arithmetic circuits, we propose a new gate structure, \textit{Compound Gates} which allows further parallel operations among encrypted bits. These gates are a hybrid of two gates, which takes two 1-bit inputs as an ordinary boolean gate but gives two different outputs. The motivation behind this novel gate structure comes from the addition circuit. For $R = A + B$, we compute $r_{i}$ and $c_i$ with the following equations:
        \begin{align}
            \label{eqn:addres}
            r_{i}&= a_{i}\oplus b_{i} \oplus c_{i-1}\\
            \label{eqn:addcarry}
            c_{i} &= {a_i}\land {b_i}\: | \:(a_{i} \oplus b_{i})\land c_{i-1}
        \end{align}
        Here, $r_i,a_i,b_i,$ and $c_i$ denotes $i^{th}$-bit of $R,A,B,$ and the carry, respectively.
        Figure~\ref{fig:bitAddition} illustrates this computation for an $n$-bit addition.

        While computing the equations~\ref{eqn:addres}, and~\ref{eqn:addcarry}, we observe that \texttt{AND} (\AndGate{}{}) and \texttt{XOR} (\XorGate{}{}) are computed on the same input bits. As these operations are independent, they can be combined into a single gate, which then can be computed in parallel. We name these gates as \textit{compound gates}. \color{black} Thus, $ a\:\Oplus\:b$ and $ a \land b$ from Equation \ref{eqn:addres} and~\ref{eqn:addcarry} can be computed as,
            $$\:\:s,c\:\: = \underbrace{a\:\Oplus\:b,a \land b}_{CONCAT}$$
            
            \noindent Here, the outputs of $s=$\AndGate{a}{b} and $c=$\XorGate{a}{b} are concatenated. The compound gates construction is analogous to the task-level parallelism in CPU, 
            where one thread performs \AndGate{}{}, while another thread performs \XorGate{}{}. 
            
            In GPU $\cparallel$, the compound gates operations are flexible as \AndGate{}{} or \XorGate{}{} can be replaced with any other logic gates. Furthermore, the structure is extensible up to $n$-bits input and $2n$-bits output.
            
    \subsection{Algebraic Circuits on GPU} \label{sec:gpu_addition}
        \subsubsection{Addition}\label{sec:gpuPar:Addition}
            
            \noindent\textit{Bitwise Addition (GPU$_1$):}
                From the addition circuit in Section~\ref{sec:openMPFra:Addition}, we did not find any data-level parallelism. However, we noticed the presence of task-level parallelism for \texttt{AND} and  \texttt{XOR}  as mentioned in the compound gates construction. Hence, we incorporated the compound gates to construct the bitwise addition circuit. 
                We also implemented the vector addition circuits using GPU$_1$ to support complex circuits such as multiplications (Section \ref{sec:gpuPar:Multiplication}).

            \noindent\textit{Number-wise Addition (GPU$_n$):}
                We consider another addition technique to benefit from bit coalescing. Here, we operate on all $n$-bits together. For $R = A + B$, we first store A in R ($R=A$). Then we compute, $Carry = $\AndGate{R}{B}, $R =$ \XorGate{R}{B}, and $\:B = Carry \ll 1$, for $n$ times.


                
                Here, we utilize compound gates to perform \AndGate{R}{B} and \XorGate{R}{B} in parallel. Thus, in each iteration, the input becomes two $n$-bit numbers, while in bitwise computation the input was two single bits.
                On the contrary, even after using compound gates, the bitwise addition (Equations~\ref{eqn:addres} and \ref{eqn:addcarry}) has more sequential blocks ($3$) than the number-wise addition ($0$). We analyze both in Section~\ref{sec:expana:additionAna}.
                
                

        \subsubsection{Multiplication}\label{sec:gpuPar:Multiplication}
            
            \textit{Naive Approach:}\label{sec:gpuPar:naiveMultiplication}
                \textcolor{black}{According to Section \ref{sec:multiplicationNaive}, multiplications have \AndGate{}{} and $\ll$ operations which can be executed in parallel. It will result in $n$-numbers where each number will have $[n,2n]$-bits due to the $\ll$. We need to accumulate these uneven sized numbers which cannot be distributed among the GPU threads.} Furthermore, the addition presents another sequential bottleneck  while adding and storing ($+=$) the results in the same memory location. Therefore, this serial addition will increase the execution  time. In the framework, we optimize the operation by introducing a tree-based approach.
                
\begin{figure}[t]
            \centering
                \includegraphics[width=0.75\linewidth, scale=0.75]{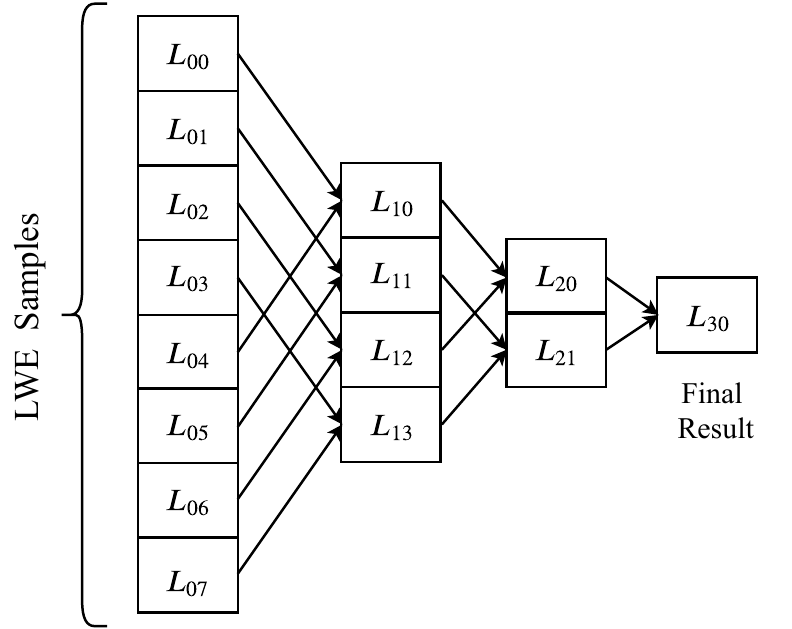}
                \caption{Accumulating $n=8$ LWE samples ($L_{ij}$) in parallel using a tree-based reduction}
                \label{fig:treemul}
            \end{figure}

                \textcolor{black}{In this approach, we divide $n$-numbers (LWE vectors) into two $n/2$ vectors. This two $n/2$ vectors are added in parallel. We repeat the process as we divide the resultant vectors int two $n/4$ vectors and add them in parallel. The process continues  until we get the final result. Notably, the tree-based approach requires $\log n$ steps for the accumulation.}
                In Figure~\ref{fig:treemul} for $n=8$, all the ciphertexts underwent \AndGate{}{} and $\ll$ in parallel, and waited for addition. Here, $L_{ij}$ represents the LWE samples (encrypted numbers), $i$ is the level, and $j$ denotes the position.
            
                Likewise vector additions, we integrated vector multiplications in similar fashion on our framework. Interestingly, we used both vector additions and multiplications in Karatsuba's algorithm which we describe next.

        \textit{Karatsuba Multiplication:}\label{sec:gpuPar:karatsuba}
                \textcolor{black}{We used Karatsuba's algorithm with some modifications in our framework to achieve further efficiency while performing multiplications. However, this algorithm requires both addition and multiplication vector operations which  tested the efficacy of these components as well.} We modified the original Algorithm~\ref{algo:karatsuba} to introduce the vector operations and rewrite the computations in Line~\ref{ln:karat:changeStart}-\ref{ln:karat:changeEnd} as:
                \begin{align*}
                \langle Temp_0,\:Temp_1 \rangle &= \langle X_0,\:X_1\rangle + \langle Y_0,\: Y_1\rangle \\
                \langle Z_0,\:Z_1,\:Z_2\rangle &=\langle X_0,\:X_1,\:Temp_0\rangle \cdot\langle Y_0,\:Y_1,\:Temp_1\rangle \\
                \langle Temp_0,\:Temp_1\rangle &= \langle Z_2,\:Z_1\rangle + \langle1,\ Z_0\rangle \\
                Z_2 &= Temp_0 + (Temp_1)'
                    \end{align*}
                
                \noindent In the above equations, $X_0,\:X_1,\:Y_0,\:Y_1,\:Z_0,\:Z_1,$ and $Z_2$ are taken from the algorithm. $\langle\hdots\rangle$ and $\cdot$ are used to denote concatenated vectors and dot product, respectively. For example, in the first equation, $Temp_0$ and $Temp_1$ store the addition of $X_0,\:Y_0$ and $X_1,\:Y_1$. 
                It is noteworthy that in the CPU $\cparallel{}$ framework, we utilized task-level parallelism to perform these vector operations as described in Section \ref{sec:openMPFra}.

        \subsubsection{Vector and Matrix Operations}\label{sec:gpupar:matrix:operations}
        
            \hfill \break
            \indent\textit{Addition:}\label{sec:gpupar:matrix:addition}
                The vector addition is a pointwise addition of the elements at their respective position. The underlying addition operation incorporates bitwise addition. Since the operation propagates bit by bit, we combine the bit from the numbers (to be added) and compute them in parallel. For example, in a vector addition of length $\ell$, we combine all the bits for the required bit position and compute the result. 

                Matrix addition also performs pointwise additions between the matrix elements. Hence, matrices represented in a row-major vector format corresponds to a vector addition operation. Therefore, we simply convert the matrices into row-major and add them utilizing the parallel vector addition. 
            \pgfplotstableread[row sep=\\,col sep=comma]{
    Bit Size,	Sequential,	OpenMP,	GPU\\
    4,	0.220166667,	0.0849522,	0.022492175\\
    8,	0.440458667,	0.167316,	0.030505775\\
    12,	0.541715,	0.220751333,	0.039014538\\
    16,	0.699724,	0.254014,	0.043097025\\
    20,	0.876932333,	0.311962333,	0.045385443\\
    24,	1.049113333,	0.370701667,	0.056019329\\
    28,	1.226243333,	0.447094667,	0.060821014\\
    32,	1.401103333,	0.507063,	0.070283043\\
}\gData

\pgfplotstableread[row sep=\\,col sep=comma]{
Bit Size,   GPU,	cuFHE,	NuFHE\\
4,  0.022492175,    0.0143494,  0.324989939\\
8,  0.030505775,    0.014048,   0.325189066\\
16, 0.043097025,   0.0143081,   0.32633481\\
32, 0.070283043,   0.029123733,   0.361531591\\
64, 0.139302,   0.0547082,   0.420085239\\
128,    0.260028,   0.094052367,   0.519757652\\
256,    0.483837,   0.169277,   0.713979673\\
512,    0.926149,   0.370075333,   1.131093264\\
1024,   1.82077,   0.674558,   2.842588472\\
}\gDataBenchmark

\pgfplotstableread[row sep=\\,col sep=comma]{
    Category,	2-Single gate,	Compound gate\\
    1,	0.0416182,	0.02103518\\
    4,	0.04498435,	0.03039582\\
    8,	0.06101155,	0.04113762\\
    16,	0.08619405,	0.069271\\
    24,	0.112038657,	0.106268\\
    32,	0.140566086,	0.138837\\
}\gDataCompound

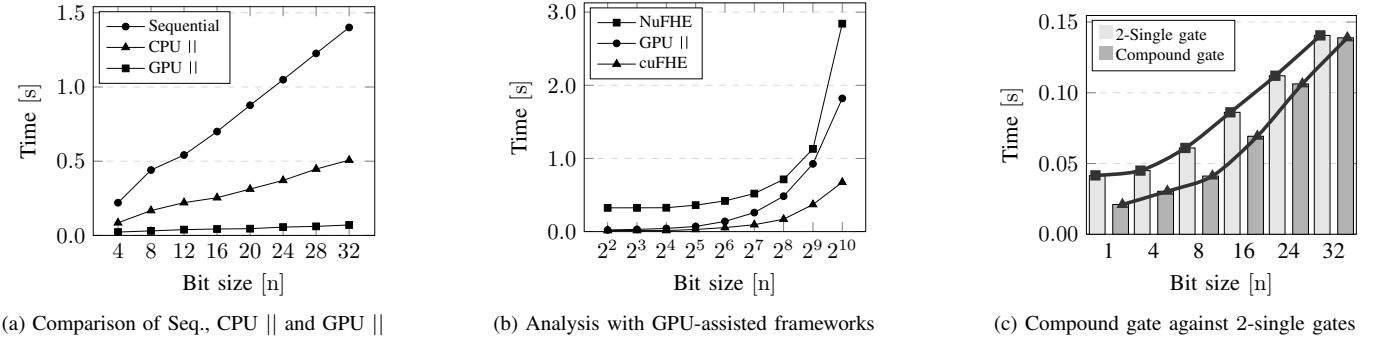
\begin{figure*}[ht]
\centering
\subfloat[Comparison of Seq., CPU $||$ and GPU $||$]{\label{fig:bitCoalescingAnalysis}{

\begin{tikzpicture}[scale=0.85]
        \begin{axis}[
                width=0.33\linewidth, 
                ymajorgrids,
                grid style={dashed,gray!30},
                xlabel= Bit size, 
                ylabel= Time,
                x unit=\si{n}, 
                y unit=\si{\second},
                xmin = 1,
                xtick=data,
                ymin = 0,
                yticklabel style={
                    /pgf/number format/fixed,
                    /pgf/number format/precision=1,
                    /pgf/number format/fixed zerofill
                },
                legend style={at={(0.02,0.98), },anchor=north west, nodes={scale=0.75, transform shape}},
                legend entries={Sequential, CPU $\cparallel$, GPU $\cparallel$},
                legend cell align={left},
                cycle list = {black,black,black}
            ]
 
                \addplot +[mark=*,mark options={scale=1}, mark size=1.5] table[x=Bit Size, y=Sequential] {\gData};
                
                \addplot +[mark=triangle*,mark options={scale=1}] table[x=Bit Size, y=OpenMP] {\gData};
                
                \addplot +[mark=square*,mark options={scale=1},mark size=1.5] table[x=Bit Size, y=GPU]{\gData};
            \end{axis}
    \end{tikzpicture}

}}\hfill
\subfloat[Analysis with GPU-assisted frameworks]{\label{fig:bitCoalescingBenchMark}{

    \begin{tikzpicture}[scale=0.85]
        \begin{axis}[
            width=0.33\linewidth,
            ymajorgrids,
            grid style={dashed,gray!30},
            xlabel= Bit size, 
            ylabel= Time,
            x unit=\si{n}, 
            y unit=\si{\second},
            xmode=log,
            log basis x={2},
            xtick=data,
            ymin = 0,
            yticklabel style={
                /pgf/number format/fixed,
                /pgf/number format/precision=1,
                /pgf/number format/fixed zerofill
            },
            legend style={at={(0.02,0.98)},anchor=north west, nodes={scale=0.75, transform shape}},
            legend entries={NuFHE, GPU $\cparallel$, cuFHE},
            legend cell align={left},
            cycle list = {black,black,black}
        ]
            \addplot +[mark=square*,mark options={scale=1}, mark size=1.5] table[x=Bit Size, y=NuFHE]{\gDataBenchmark};
            \addplot +[mark=*,mark options={scale=1}, mark size=1.5] table[x=Bit Size, y=GPU] {\gDataBenchmark};
            
            \addplot +[mark=triangle*] table[x=Bit Size, y=cuFHE] {\gDataBenchmark};
        \end{axis}
    \end{tikzpicture}
}}\hfill
\subfloat[Compound gate against $2$-single gates]{\label{fig:compoundGateAna}{
\begin{tikzpicture}[scale=0.85]
            \begin{axis}[
                width=0.32\linewidth,
                ybar=.12cm,
                symbolic x coords={1, 4, 8, 16, 24, 32},
                xtick=data,
                enlarge x limits=0.1,
                xtick align=inside,
                xtick style={draw=none},
                xlabel=Bit size,
                x unit=\si{n},
                bar width=7pt,
                ymin=0,
                yticklabel style={
                    /pgf/number format/fixed,
                    /pgf/number format/precision=2,
                    /pgf/number format/fixed zerofill
                },
                ymajorgrids = true,
                grid style={dashed,gray!30},
                ylabel= Time,
                y unit=\si{\second},
                legend image code/.code={
                    \draw[#1,draw=none,/tikz/.cd,yshift=-0.25em]
                        (0cm,1pt) rectangle (6pt,7pt);
                        },
                legend style={at={(0.02,0.98)},anchor=north west, nodes={scale=0.75, transform shape}},
                legend cell align={left},
                no marks,
                every axis plot/.append style={fill, fill opacity=0.6},
                every axis plot/.append style={line width=0.2pt}
            ]
                \addplot [draw=black, fill=black!10] table[x=Category, y=2-Single gate] {\gDataCompound};
                \addlegendentryexpanded{2-Single gate}
                \addplot [draw=black, fill=black!30] table[x=Category, y=Compound gate] {\gDataCompound};
                \addlegendentryexpanded{Compound gate}
        \end{axis}
        
        \begin{axis}[
                    hide axis,
                    symbolic x coords={1, 4, 8, 16, 24, 32},
                    enlarge x limits=0.1,
                    xtick align=inside,
                    xtick style={draw=none},
                    width=0.32\linewidth,
                    ymin=0,
                    cycle list = {black!80,black!80},
                ]
            \addplot+[shift={(-6pt, 0pt)},ultra thick,smooth, mark=square*, mark size=1.5] table[x=Category, y=2-Single gate] {\gDataCompound};
            \addplot+[shift={(6pt, 0pt)},ultra thick,smooth, mark=triangle*, mark size=1.5] table[x=Category, y=Compound gate] {\gDataCompound};
        \end{axis}
    \end{tikzpicture}

}}

\caption{Performance analysis of GPU-accelerated TFHE with the sequential and CPU $||$ frameworks (\ref{fig:bitCoalescingAnalysis}), and comparison with the existing GPU-assisted libraries (\ref{fig:bitCoalescingBenchMark}). Figure~\ref{fig:compoundGateAna} presents the performance of compound gates against 2-single gate operations}
\end{figure*}

            \textit{Multiplication:}\label{sec:gpupar:matrix:multiplication}
            Analogous to additions, vector multiplications are also a pointwise operation. Here, we compute the \texttt{AND} and  \texttt{left shift} operations in parallel and accumulate the values in a tree-based approach as described in regular multiplications (Section \ref{sec:gpuPar:Multiplication}).
            
            \textcolor{black}{Unlike addition, matrix multiplication is more complicated as it requires more computations. Section~\ref{sec:openMPFra:matrix:multiplication} presents a schematic computation for a $2\times 2$ matrix along with the possible parallel computations. Notably, for two $n$-ranked \textit{squared} (for brevity)  matrix multiplication, we require $n^{3}$  multiplication operations. Here,  we separate all the multipliers and multiplicands into two vectors and perform parallel vector operations.} For example,  in a square matrix of rank $16$, each vector length for multiplication will be $4096$. Furthermore, for each $16$-bit vector components (matrix elements), the computation raises to $4096\times 16 \times 16$-bits computation for parallel multiplication.
            
            \color{black}
            Therefore, multiplying these vectors require larger GPU memory which we wanted to avoid. It also required a large number of threads for computing and storing the LWE samples as well. Although, the threads can be reused sequentially, this problem is more severe due to the fixed GPU memory constraint. Hence, matrix multiplications on larger dimensions can essentially run out of GPU memory.
            

            Therefore, we consider a different technique named Cannon's Algorithm~\cite{cannon1969cellular}. 
            The algorithm consists of cycles containing multiplication, addition, and shifting the matrix elements. Each cycle also depends on the values generated by the previous cycle, incurring a sequential bottleneck. Nevertheless, the algorithm provides a much needed scalability for large-scale multiplication instead of using a considerable amount of GPU memory which is often not present.
            

            In each cycle, we first perform the multiplication using the vector operations. Then, we add on the multiplied data using the parallel vector additions. Lastly, we shift the elements' positions for the next round.

\color{black}

\section{Experimental Analysis}\label{sec:expAnalysis}
    
    The experimental environment included an Intel(R) Core\texttrademark{}  i7-2600 CPU having 16 GB system memory with a  NVIDIA GeForce GTX 1080 GPU with 8 GB memory~\cite{gpuspec}. The CPU and GPU contained $8$ and $40,960$ hardware threads, respectively. We used the same setup to analyze all  three frameworks: sequential, CPU $\cparallel{}$, GPU $\cparallel{}$. 
    
    \color{black}
    We use two metrics for the comparison: a) execution time and b) speedup $=\frac{T_{seq}}{T_{par}}$.
    Here, $T_{seq}$ and $T_{par}$ are the time for computing the sequential and the parallel algorithm. In the following sections, we gradually analyze the complicated arithmetic circuits using the best results from the foregoing analysis. 
    
    \subsection{GPU-accelerated TFHE}\label{sec:expAna:GPUtfhe}
        \begin{table*}[t]
    \centering
    \caption{Computation time (ms) for Bootstrapping, Key Switching and Misc. for sequential and GPU framework}
    \label{tbl:differentialtime}
    \begingroup
    \begin{tabular}{c|c|c|c|c|c|c|c|c|c}
        \hline
          Bit Size  & \multicolumn{4}{c|}{\textbf{Sequential}} &   & \multicolumn{4}{c}{\textbf{GPU}}   \\ \hline
        $\textbf{n}$ &   \textbf{Bootstrapping}   &   \textbf{Key Switch}  &   \textbf{Misc.}   &   \textbf{Total}   &   &   \textbf{Bootstrapping}   &   \textbf{Key Switch}  &   \textbf{Misc.}   &   \textbf{Total}   \\ \hline
        2	&   68.89           &	17.13	    &   27.04	&   113.05	&   &   19.64   &   2.65    &   0.45    &	22.74\\ \hline
        4	&   138.02          &	34.18	    &   47.97	&   220.17	&   &   18.86   &	2.69    &	0.08    &	21.63\\ \hline
        8	&   275.67          &	68.31	    &   96.48	&   440.46	&   &   27.83   &	2.69    &	0.06    &	30.58   \\ \hline
        16	&   137.25          &	137.25	    &   425.22	&   699.72	&   &   40.70    &	2.91    &	0.44    &	44.06\\ \hline
        32	&   274.3           &	274.30	    &   852.51	&   1401.10	&   &   66.74   &	3.34    &	0.42    &	70.50\\ \hline
    \end{tabular}
    \endgroup
\end{table*}
        Initially, we discuss our performance over boolean gate operations, which is deemed as a building blocks of any computation. Figure~\ref{fig:bitCoalescingAnalysis} depicts the execution time difference among the sequential, CPU $\cparallel{}$ and GPU $\cparallel{}$ framework  for $[4, 32]$-bits. The sequential \texttt{AND} operation takes a minimum of 0.22s ($4$-bit) while the runtime increases to 1.4s for $32$-bits.
        
        \color{black}
        In the GPU $\cparallel$ framework, bit coalescing facilitates storing LWE samples in contiguous memory and takes advantage of available vector operations. Thus, it helps to reduce the execution time from $0.22-1.4s$ to $0.02-0.06s$ for $4$ to $32$-bits. Here, for $32$-bits, our techniques provide a $20\times$ speedup. Similar improvement is foreseen in the CPU $\cparallel$ framework as we divide the number of bits by the available threads. However, the execution time increases for CPU framework since there is only a limited number of available threads. \textcolor{black}{This limited number of threads is one of the primary motivations behind utilizing GPU.}

        Then, we further scrutinize the execution time by dividing gate operations into three major components---\begin{enumerate*}[{label=\textit{\alph*)}}]\item Bootstrapping, \item Key Switching, and \item Miscellaneous.\end{enumerate*} We selected the first two as they are the most time-consuming operations and fairly generalizable to other HE schemes. 
        Table~\ref{tbl:differentialtime} shows the difference in execution time between the sequential and the GPU $\cparallel$ for $\{2,\ldots,32\}$-bits. \textcolor{black}{We show that the execution time increment is less compared to the sequential approach.} 

        \textcolor{black}{We further investigated the bootstrapping performance in GPU $\cparallel$ framework for the boolean gate operations. Our cuda enabled FFT library takes the LWE samples in batches and performs the FFT in parallel. However, due to the h/w limitations, the number of batches to be executed in parallel is limited. It can only operate on a certain number of batches at once and next batches are kept in a queue. Hence, a sequential overhead occurs for a large number of batches that can increase the execution time.}

        \textcolor{black}{Under the same h/w setting, we benchmark our proposed framework with the existing GPU-based libraries (cuFHE and NuFHE).} Although our GPU $\cparallel$ framework outperforms NuFHE for different bit sizes (Figure~\ref{fig:bitCoalescingBenchMark}), the performance degrades for larger bit sizes w.r.t. cuFHE. As the cuFHE implementation focuses more on the gate level optimization, we focus  on the arithmetic circuit computations. \textcolor{black}{In Section \ref{sec:expana:additionAna}, we analyze our arithmetic circuits where our framework outperforms the existing GPU libraries.}
    
    \subsection{Compound Gate Analysis}\label{sec:expAna:compoundgateAna}
    According to Section~\ref{sec:gpuPar:compound},  the compound gates are used to improve the execution time for additions or multiplications. \textcolor{black}{Since, the existing frameworks do not provide these optimizations, we benchmark the compound gates with the proposed single gate computations. Figure~\ref{fig:compoundGateAna} illustrates the performance of one compound gate over $2$-single gates computed sequentially. We performed several iterations for different number of bits ($1,\ldots,32$) as shown on the X-axis while the Y-axis represents the execution time. Notably,a 32-bit compound gates will have two $32$-bit inputs and output two $32$-bits.}

    \textcolor{black}{Here, bit coalescing improves the execution time as it takes only 0.02s for one compound gates evaluation, compared to 0.04s on performing 2-single gates sequentially.} However, Figure~\ref{fig:compoundGateAna} shows an interesting trend in the execution time between $2$-single gates and one compound gates evaluation. The gap favoring the compound ones tends to get narrower for higher number of bits. For example, the speedup for 1-bit happens to be ${0.04}/{0.02}=2$ times whereas it reduces to $1.01$ for 32-bits. \textcolor{black}{The reason behind this diminishing performance is the \textit{asynchronous launch queue} of GPUs.}
        
    \textcolor{black}{As mentioned in Section~\ref{sec:gpuPar:TfheCons}, we use batch execution for the FFT operations. Hence, the number of parallel batches depends on the asynchronous launch queue size of the underlying GPU which can delay the FFT operations for a large number batches. This ultimately adversely affects the speedup for large LWE sample vectors. Nevertheless, the analysis shows that the 1-bit compound gates is the most efficient, and we employ it in the following arithmetic operations.}

    \subsection{Addition}\label{sec:expana:additionAna}

        \begin{table}
    \centering
    
    \caption{Execution time (sec) for the $n$-bit addition}
    \label{tbl:additionComparison}
    \begingroup
    \begin{tabular}{l|c|c|c}
        \hline
        \textbf{Frameworks}                 & \textbf{16-bit} & \textbf{24-bit} & \textbf{32-bit} \\ \hline
        Sequential                          & 3.51            & 5.23            & 7.04            \\ \hline
        cuFHE \cite{cuFHE}                              & 1.00            & 1.51            & 2.03            \\ \hline
        NuFHE \cite{NuFHE}                              & 2.92            & 3.56            & 4.16            \\ \hline
        Cingulata  \cite{cingulata}                         & 1.10            & 1.63            & 2.16            \\ \hline
        \multicolumn{4}{c}{\textbf{Our Methods}}       \\
        \hline
        CPU $\cparallel$                              & 3.51            & 5.23            & 7.04            \\ \hline
        GPU$_n\cparallel$                   & 0.94            & 2.55            & 4.44            \\ \hline
        GPU$_1\cparallel$                   & 0.98            & 1.47            & 1.99            \\ \hline
        
    \end{tabular}
    \color{black}
    \endgroup
\end{table}
        Table~\ref{tbl:additionComparison} presents a comparative analysis of the addition operation for 16, 24, 32-bit encrypted numbers. We consider our proposed frameworks: sequential, CPU $\cparallel$, and GPU $\cparallel{}$,
        and benchmark them with cuFHE \cite{cuFHE}, NuFHE \cite{NuFHE} and Cingulata \cite{cingulata}. Furthermore, we present the performance of two variants of addition operation: GPU$_{n}\cparallel$ (number-wise) and GPU$_{1}\cparallel$ (bitwise) as discussed in Section \ref{sec:gpuPar:Addition}.
        
        Table~\ref{tbl:additionComparison} demonstrates that GPU$_n\cparallel$ performs better than the sequential and CPU $\cparallel{}$ circuits. The GPU$_n$ provides a  $3.72\times$ speedup for $16$-bits whereas $1.58\times$ for $32$-bit. However,  GPU$_n\cparallel$ performs better only for $16$-bit additions compared to GPU$_1\cparallel$.
        For $24$ and $32$-bit additions, GPU$_{1}\cparallel$ performs around $2\times$ better than GPU$_n\cparallel$. \textcolor{black}{This improvement in essential as it reveals the algorithm to choose between  GPU$_{1}\cparallel$ and  GPU$_{n}\cparallel$.}

        \begin{table}[t]
\centering
\caption{Execution time (sec) for vector addition}
\label{tbl:vecAdditionTime}
\resizebox{.47\textwidth}{!}{\begin{tabular}{c|c|c|c|c|c|c|c}
\hline
Length& \multicolumn{3}{c|}{\textbf{16-bit}} &  & \multicolumn{3}{c}{\textbf{32-bit}} \\ \hline
$\ell$  &   \textbf{Seq.} &   \textbf{CPU} $\cparallel$ &   \textbf{GPU} $\cparallel$ &   &    \textbf{Seq.} &   \textbf{CPU} $\cparallel$ &   \textbf{GPU} $\cparallel$ \\ \hline
4	    &       13.98	&       5.07	    &       1.27	    &   &       28.05	&       10.02	    &       2.56    \\ \hline
8	    &       27.86	&       9.96	    &       1.78	    &   &       56.01	&       19.29	    &       3.58    \\ \hline
16	    &       55.66	&       19.65	    &       2.82	    &   &       111.3	&       38.77	    &       5.70    \\ \hline
32	    &       111.32	&       38.99	    &       5.41	    &   &       224.31	&       77.18	    &       11.22    \\ \hline
\end{tabular}}
\end{table}
        
        Although, both addition operations (GPU$_n\cparallel$ and GPU$_1\cparallel$) utilize compound gates, they differ in the number of input bits ($n$ and $1$ for GPU$_n\cparallel$ and GPU$_1\cparallel$, respectively). Since the compound gates performs better for smaller bits (Section~\ref{sec:expAna:compoundgateAna}), the bitwise addition performs better than the number-wise addition for 24/32-bit operations. Hence, we utilize bitwise addition for building other circuits.
        
        \textcolor{black}{NuFHE and cuFHE do not provide any arithmetic circuits in their library. Therefore, we implemented such circuits on their library and performed the same experiments. Additionally, we considered Cingulata \cite{cingulata} (a compiler toolchain for TFHE) and compared the execution time. Table~\ref{tbl:additionComparison} summarizes all the results, where we found our proposed addition circuit (GPU$_1\cparallel$) outperforms the other approaches.}
        
        We further experimented on the vector additions adopting the bitwise addition and showed the analysis in Table~\ref{tbl:vecAdditionTime}.  Like addition, the performance improvement on the vector addition is also noticeable. The framework scales by taking  similar execution time for smaller vector lengths $\ell \leq 8$. However, the execution time increases for longer vectors as they involve more parallel bit computations, and consequently, increase the batch size of FFT operations. \textcolor{black}{The difference is clearer on 32-bit vector additions with $\ell=32$ which takes almost twice the time of $\ell=16$. However, for $\ell\leq8$, the executions times are almost similar due to the parallel computations. In Section~\ref{sec:expAna:compoundgateAna} we have discussed this issue which relies on the FFT batch size.  Notably, Figure~\ref{fig:compoundGateAna} also aligns with this evidence as  the larger batch size for FFT on GPUs affects the speedup. For example, $\ell = 32$ will require more FFT batches compared to $\ell = 16$ which requires more time to finish the addition operation.}
        We did not include other frameworks in Table~\ref{tbl:vecAdditionTime}, since our GPU $\cparallel$ performed better comparing to the others in Table~\ref{tbl:additionComparison}. 
        
        
    \subsection{Multiplication}
        \begin{table}[t]
    \centering
    \caption{Multiplication execution time (sec) comparison}
    \label{tbl:multiplicationComparison}
    \begin{tabular}{l|c|c|c}
        \hline
        \textbf{Frameworks} & \textbf{16-bit} & \textbf{24-bit} & \textbf{32-bit} \\ \hline
        \multicolumn{4}{c}{\textbf{Naive}}                                      \\ \hline
        Sequential          & 120.64          & 273.82          & 489.94          \\ \hline
        CPU $\cparallel$     & 52.77           & 101.22          & 174.54          \\ \hline
        GPU $\cparallel$     & 11.16           & 22.08           & 33.99           \\ \hline
        cuFHE~\cite{cuFHE}               & 32.75           & 74.21           & 132.23          \\ \hline
        NuFHE~\cite{NuFHE}               & 47.72           & 105.48          & 186.00          \\ \hline
        Cingulata~\cite{cingulata}           & 11.50           & 27.04           & 50.69           \\ \hline
        \multicolumn{4}{c}{\textbf{Karatsuba}}                                  \\ \hline
        CPU $\cparallel$     & 54.76           & -               & 177.04          \\ \hline
        GPU $\cparallel$     & 7.6708          & -               & 24.62           \\ \hline
    \end{tabular}
\end{table}

        \begin{table}[t]
\centering
\caption{Execution time (min) for vector multiplication}
\label{tbl:vecMultiplicationTime}
\resizebox{\linewidth}{!}{
\begingroup
\begin{tabular}{c|c|c|c|c|c|c|c}
\hline
    Length   & \multicolumn{3}{c|}{\textbf{16-bit}} &  & \multicolumn{3}{c}{\textbf{32-bit}} \\ \hline
$\ell$& \textbf{Seq.} & \textbf{CPU} $\cparallel$ & \textbf{GPU} $\cparallel$   &  & \textbf{Seq.} & \textbf{CPU} $\cparallel$ & \textbf{GPU} $\cparallel$   \\ \hline
4       &   8.13    &   3.25    &   0.41    &   &   32.56       &   12.15   &   1.61    \\ \hline
8       &   16.29   &   6.17    &   0.75    &   &   65.12       &   23.48   &   2.96    \\ \hline
16      &   32.62   &   11.93   &   1.40    &   &   130.31      &   46.39   &   5.62    \\ \hline
32      &   65.15   &   23.58   &   2.68    &   &   260.52      &   92.44   &   10.79\\ \hline
\end{tabular}
\endgroup
}
\end{table}
        
        The multiplication operation uses a sequential accumulation (reduce by addition) operation. Instead, we use a tree-based vector addition approach (discussed in Section~\ref{sec:gpuPar:Multiplication}) and gain a significant speedup. Table~\ref{tbl:multiplicationComparison} portrays the execution times for the multiplication operations using the frameworks.
        Here, we employed all available threads on the machine. Like the addition circuit performance, here GPU $\cparallel{}$ outperforms the sequential circuits and CPU $\cparallel{}$ operations by a factor of $\approx 11$ and $\approx 14.5$, respectively for $32$-bit multiplication.
    
        We further implemented the multiplication circuit on cuFHE and NuFHE. Table~\ref{tbl:multiplicationComparison} summarizes the results comparing our proposed framework with cuFHE, NuFHE, and Cingulata. Our GPU $\cparallel{}$ framework is faster in execution time than the other techniques.  Notably,  the  performance improvement is scalable with the increasing number of bits. This is due to tree-based additions following the reduction operations and computing all boolean gate operations by coalescing the bits altogether.
        
        Besides, we also analyze vector multiplications available in our framework and present a comparison among the frameworks in Table~\ref{tbl:vecMultiplicationTime}. We found out an increase in execution time for a certain length (\eg, $\ell=32$ on $16$-bit or $\ell=4$ on $32$-bit), which is similar to the issue in vector addition (Section~\ref{sec:expana:additionAna}).
        Hence, the vector operations from $\ell\leq16$ can be sequentially added to compute arbitrary vector operations. For example, we can use two $\ell=16$ vector multiplication to compute  $\ell=32$ multiplication resulting around $11$ mins. In the vector analysis, we did not add the computations over the other frameworks since our
        framework surpassed their achievements for a single multiplications. 
        
    \subsection{Karatsuba Multiplication}\label{sec:karat_analysis}
        In Table~\ref{tbl:multiplicationComparison}, we provide execution time for $16$ and $24$-bit Karatsuba multiplication over encrypted numbers as well. In the CPU $\cparallel$ construction of the algorithm, the execution time does not improvement, rather it increases slightly.
        We observed that for both 16 and 32-bit multiplication, Karatsuba outperforms naive GPU$\cparallel{}$ multiplication algorithm on GPU by $1.50$ times. 
        Karatsuba multiplication can also be considered a complex 
        arithmetic operation as it comprises of both addition, multiplication, and vector operations. However, the CPU $\cparallel{}$ framework did not provide such difference in performance as it took more time for the \texttt{fork-and-join} threads required by the divide and conquer algorithm.
        
        
    \subsection{Matrix Operations}\label{sec:analysis:matrix:multiplication}

        \begin{table}[t]
    \centering
    \caption{Matrix multiplication execution time (min)}
    \label{tbl:matmulana}
    \begingroup
    \begin{tabular}{c|c|c|c}
        \hline
        \textbf{Dimension} & \textbf{Sequential} & \textbf{CPU} $\cparallel$ & \textbf{GPU} $\cparallel$ \\ \hline
        $2 \times 2$   & 17.07          & 10.62      & 0.86 \\ \hline
        $4 \times 4$   & 136.68         & 47.78      & 5.90 \\ \hline
        $8 \times 8$   & 1,090.12          & 351.82      & 43.95 \\ \hline
        $16 \times 16$  & 8,717.89          & 2,514.34      & 186.23  \\ \hline
    \end{tabular}
    \endgroup
\end{table}
        From Section~\ref{sec:gpupar:matrix:operations}, it is evident that the vector addition  represents matrix additions as both operations are done point-wise. Therefore, Table~\ref{tbl:vecAdditionTime} can be extended to represent the execution time for the matrix additions, where $\ell$ becomes the number of elements of the matrices.
        Table~\ref{tbl:matmulana} enlists the matrix multiplication execution time  for different dimensions using Cannon's algorithm~\cite{cannon1969cellular}. For a $16\times16$ matrix, GPU $\cparallel{}$ achieves a $\approx 48$ and $\approx 15$ times speedup compared to the  sequential and CPU $\cparallel{}$ approach, respectively.

    \color{black}    
    \begin{table}[t]
\centering
\caption{Execution time (min) for Linear Regression}
\label{tbl:linearRegression}
\begin{tabular}{l|c|c|c|c}
\hline
\textbf{Datasets}                  & \textbf{\#Rows}      & \textbf{\#Attributes}  & \textbf{Data Type} & \textbf{Time} \\ \hline
\multirow{2}{*}{Dataset 1} & \multirow{4}{*}{200} & \multirow{2}{*}{10} & Numerical          & 163.38        \\ \cline{4-5} 
                           &                      &                     & Binary             & 53.91         \\ \cline{1-1} \cline{3-5}
\multirow{2}{*}{Dataset 2} &  & \multirow{2}{*}{20} & Numerical          & 268.86        \\ \cline{4-5} 
                           &                      &                     & Binary             & 67.88         \\ \hline
\multirow{2}{*}{Dataset 3} & \multirow{4}{*}{300} & \multirow{2}{*}{10} & Numerical          & 245.38        \\ \cline{4-5} 
                           &                      &                     & Binary             & 80.91         \\ \cline{1-1} \cline{3-5}
\multirow{2}{*}{Dataset 4} &  & \multirow{2}{*}{20} & Numerical          & 403.85        \\ \cline{4-5} 
                           &                      &                     & Binary             & 95.88         \\ \hline
\end{tabular}
\end{table}

    \subsection{Application: Linear Regression}
        We employed the arithmetic operations (vector/matrix addition, multiplication) to compute linear regression models as an application to test the efficacy of the framework. We produced four synthetic dataset with different number of instances (rows) and attributes (columns), and tabulated the execution times in Table~\ref{tbl:linearRegression}. Each datasets had two variants: binary and numeric values. As the multiplications are essentially  \texttt{AND} operations for binary values, it takes less time compared to the numeric dataset.
    \color{black}

\section{Discussion}
    In this section, we provide answers to the following questions about our proposed framework:
    

    \noindent \textbf{Is the proposed framework sufficient to implement any computations?}
    In this article, we show how to implement boolean gates properly using GPUs to gain performance improvement. We then show how to compute addition, multiplication, and matrix operations using the proposed framework. Implementing more complex algorithms such as \textit{secure machine learning} \cite{xie2014crypto,takabi2016privacy} are beyond the scope of this paper.  In future work, we will investigate how to further optimize the framework for machine learning algorithms.  
    Note that we have implemented a FHE scheme. Hence,  any computable function can be implemented using our framework. 

    \noindent \textbf{For GPU $\cparallel{}$ framework, how do we compute on encrypted data larger than the fixed GPU memory?} The fixed GPU memories and their variations in access speeds are limitations for any GPU $\cparallel{}$ application. Similar problems also occur in deep learning while handling larger datasets. The solution includes batching the data or using multiple GPUs. Our proposed framework can also avail such solutions as it can easily be extended to accommodate larger ciphertexts.

    \noindent \textbf{How can we achieve further speedup on both frameworks?} On the CPU $\cparallel{}$ framework, we have attempted most H/W or S/W level optimizations to the best of our knowledge. However, our GPU $\cparallel{}$ framework partially relied on the global GPU memory, which is slower than its counterparts. This is critical as different device memories offer variant read/write speeds. Notably, shared memory (\texttt{L1}) is the fastest memory after register.
    Our implementation uses a combination of shared and global memory due to the ciphertext size. In the future, we would like to utilize only the shared memory, which is much smaller but should provide better speedup compared to the current approach.

    \noindent \textbf{How the bit security level would affect the reported speedup?} The current framework is analogous to the existing implementation of TFHE~\cite{TFHEcode} providing $110$-bit security which might not be sufficient for some applications. However, our GPU $\cparallel$ framework can accommodate any change for the desired bit security level. Nevertheless, such change will  change the execution times as well. For example, any less security level than $110$-bits will result in faster execution and likewise for a higher bit security. We will include and analyze the speedup for the dynamic bit security levels in future. 
    
    

\section{Related Works}\label{sec:relWorls}
    \textcolor{black}{In this section, we discuss the other HE schemes from Table \ref{tbl:comparativeAna} and categorize schemes based on their number representation: \begin{enumerate*}[{label=\textit{\alph*)}}] \item bit-wise, \item modular  and \item approximate. \end{enumerate*} }
    

 

    \noindent\textbf{Bitwise Encryption} usually takes the bit representation of any number and encrypts accordingly.  The computations are also done bit-wise as each bit can be considered independent from another. This bit-wise representation is crucial for our parallel framework as it offers less dependency between bits which we can operate in parallel. Furthermore, it provides faster bootstrapping and smaller ciphertext size, which can be easily tailored for the fixed memory GPUs. This concept is formalized and named as GSW \cite{gentry2013homomorphic} around 2013, and it was later improved in subsequent works \cite{ducas2014,chillotti2016faster,chillotti2017faster}. 
    
    \noindent\textbf{Modular Encryption} schemes utilize a fixed modulus $q$ which denotes the size of the ciphertexts. There have been many developments \cite{brakerski2014efficient,brakerski2014leveled} in this direction as they offer a reasonable execution time (Table~\ref{tbl:comparativeAna}). The addition and multiplication times from FV~\cite{fan2012somewhat} and SEAL \cite{sealcrypto} show the difference as they are much faster compared to our GPU-based 
    framework.  
    
    However, these schemes do a trade-off between the bootstrapping and the efficiency as they are often designated as somewhat homomorphic encryption. Here, in most cases, the number of computations or the level of multiplications are predefined as there is no procedure for noise reduction
    Decryption is performed after the desired computation. Furthermore, the encrypted data evidently suffers from larger ciphertexts as the value of $q$ is picked from large numbers.
    
    For example, we selected the ciphertext modulus of 250 and 881 bits for FV-NFLlib~\cite{fan2012somewhat} and  SEAL~\cite{sealcrypto}, respectively. The polynomial degrees  ($d$) were chosen 13 and 15 for the two frameworks as it was required to comply with the targeted bit security to populate Table~\ref{tbl:comparativeAna}. It is noteworthy that smaller $q$ and $d$ will result in faster runtime and smaller ciphertexts, but they will limit the number of computations as well. Therefore, this modular representation requires 
    to fix the number of homomorphic operations limiting the use cases.

    \noindent \textbf{Approximate Number} representations are recently proposed by Cheon \etal (CKKS \cite{cheon2017homomorphic}) in 2017. These schemes also provide efficient Single Instruction Multiple Data (SIMD)~\cite{flynn1972some} operations similar to the modular representations as mentioned above. However, they have an inexact but efficient bootstrapping mechanism which can be applied in less precision-demanding applications. The cryptosystem also incurs larger ciphertexts (7MB) similar to the modular approach as we tested it for $q = 1050$ and $d = 15$. Here, we did not discuss HELib~\cite{halevi2014algorithms}, the first cornerstone of all HE implementations since its cryptosystem BGV~\cite{brakerski2014leveled} is enhanced and utilized by the other modular HE schemes (such as SEAL~\cite{sealcrypto}).
    

    The goal of this work is to parallelize an FHE scheme. Most HE schemes that follow modular encryption are either somewhat or adopt inexact bootstrapping. Besides, their expansion after encryption requires more memory. Hence, we choose the bitwise and bootstrappable encryption scheme: TFHE.

    \noindent\textbf{Hardware Solutions} are less studied and employed to increase the efficiency of FHE computations. Since the formulation of FHE~\cite{gentry2009fully} with ideal lattices, most of the efficiency improvements are considered from the standpoint of asymptotic runtimes. A few approaches considered the incorporation of existing multiprocessors (\eg, GPU) or FPGAs~\cite{doroz2015accelerating} to achieve faster homomorphic operation. Dai and Sunar ported another scheme LTV~\cite{lopez2012fly} to GPU-based implementation \cite{dai2015cuhe,dai2015accelerating}. LTV is a variant of HE that performs a limited number of operations on a given ciphertext. 

    Lei~\etal ported  FHEW-V2~\cite{ducas2014}  to GPU~\cite{lei2019accelerating} and extended the boolean implementation to $30$-bit addition and $6$-bit multiplication with a speed up $\approx2.5$. Since TFHE extends FHEW and  performs better than it predecessor, we consider TFHE as our baseline framework.

    In 2015, a GPU based HE scheme  CuHE~\cite{dai2015cuhe} was proposed. However, it was not fully homomorphic as it did not have bootstrapping, hence we do not include it in our analyses. Later in 2018, two GPU FHE libraries cuFHE~\cite{cuFHE} and NuFHE~\cite{NuFHE} were released. Both the libraries focused on optimizing of the boolean gate operations. Recently, Yang~\etal~\cite{yang2019efficiency}  benchmarked cuFHE and its predecessor TFHE, and analyzed the speedup which we also discuss in our paper (Table \ref{tbl:comparativeAna}).
    
    \color{black}
    
    Our experimental analysis shows that only performing the boolean gates in parallel is not sufficient to reduce the execution time of higher level circuit (i.e., multiplication). Hence, besides employing GPU for homomorphic gate operations, we focus on arithmetic circuit. For example, we are 3.9 times faster than cuFHE in 32-bit multiplications. 
    
    Recently, Zhou~\etal improved TFHE by reducing and performing the serial operations of bootstrapping in parallel~\cite{zhou2018faster}. However, they did use any hardware acceleration to the existing FHE operations. We consider this work as an essential future direction that can be integrated to our framework for better executing times.
    \color{black}
    
\section*{Acknowledgments}
    We sincerely thank the reviewers for their insightful comments. This research was supported in part by the NSERC Discovery Grants (RGPIN-2015-04147). We also  acknowledge the support from NVIDIA for their GPU support and Amazon Cloud Grant.

\section{Conclusion}\label{sec:conclusion}
    In this paper, we constructed the algebraic circuits for FHE, which can be utilized by arbitrary complex operations. Furthermore, we explored the CPU-level parallelism for improving the execution time of the underlying FHE computations. Our notable contribution is the proposed GPU-level parallel framework that utilizes novel optimizations such as bit coalescing, compound gate, and tree-based vector accumulation. Experimental results show that the proposed method is $20\times$ and $14.5\times$ faster than the existing technique for computing boolean gates and multiplications respectively (Table \ref{tab:contribution}).



\bibliographystyle{IEEEtran}
\bibliography{main}

\end{document}